\documentclass{aa}

\usepackage{txfonts}
\usepackage{graphicx}

\begin{document}

\title{On the Periodicity of Oscillatory Reconnection}

\author{J.~A.~McLaughlin\inst{1} \and J.~O. Thurgood\inst{1} \and  D. MacTaggart\inst{2}}

\offprints{J.~A.~McLaughlin, \email{james.a.mclaughlin@northumbria.ac.uk}}

\institute{School of Computing, Engineering \& Information Sciences, Northumbria University, Newcastle Upon Tyne, NE1 8ST, UK \and School of Engineering, Computing \& Applied Mathematics, University of Abertay,  Dundee, DD1 1HG, UK} 

\date{Received 15 August 2012 / Accepted 31 October 2012}

\authorrunning{McLaughlin {{et al.}}}
\titlerunning{On the Periodicity of Oscillatory Reconnection}

%----------------------------------------------------------------
%----------------------------------------------------------------
%----------------------------------------------------------------

\abstract{Oscillatory reconnection is a time-dependent magnetic reconnection  mechanism that naturally  produces periodic outputs from aperiodic drivers.}
{This paper aims to quantify and measure the periodic nature of oscillatory reconnection for the first time.}
{We solve the compressible, resistive, nonlinear MHD equations using 2.5D numerical simulations.}
{We identify two distinct periodic regimes: the impulsive and stationary phases. In the impulsive phase, we find the greater the amplitude of the initial velocity driver, the longer the resultant current sheet and the earlier its formation. In the stationary phase, we  find that the  oscillations are exponentially decaying and for driving amplitudes $6.3 - 126.2\:$km/s, we measure stationary-phase periods in the range $56.3 - 78.9\:$s, i.e. these are high frequency ($0.01 - 0.02\:$Hz) oscillations. In both phases, we find that the greater the amplitude of the initial velocity driver, the shorter the resultant period, but note that different physical processes and periods are associated with both phases.}
{We conclude that the oscillatory reconnection mechanism behaves akin to a {\emph{damped harmonic oscillator}}.}

%\keywords{Magnetohydrodynamics (MHD) -- Waves -- magnetic null point -- coronal heating -- Sun:~corona -- Sun:~ magnetic fields -- Sun:~oscillations  }


\maketitle

%----------------------------------------------------------------
%----------------------------------------------------------------
%----------------------------------------------------------------
\section{Introduction}\label{section1}
%----------------------------------------------------------------
%----------------------------------------------------------------
%----------------------------------------------------------------

Traditionally, magnetic reconnection  and MHD wave theory   have been viewed as separate areas of solar physics (see, e.g., Priest \& Forbes \cite{magneticreconnection2000}; Roberts \cite{Bernie}; De Moortel \cite{DeMoortel2005};  Nakariakov \& Verwichte \cite{NV2005}; De Moortel \& Nakariakov \cite{DN2012}). However, this is a misconception: we know that (steady-state) reconnection models not only generate outflows/waves, but also require inflows/waves (e.g. Parker \cite{Parker}; Sweet \cite{Sweet}; Petschek \cite{Petschek}). Several authors  have already challenged this point-of-view (e.g. Craig \& McClymont \cite{CraigMcClymont1991}; Longcope \& Priest \cite{LP2007}; Murray et al. \cite{Murray2009}; McLaughlin et al. \cite{McLaughlin2009}; \cite{McLaughlin2012}) and their investigations contribute to our understanding of dynamic or time-dependent models of magnetic reconnection. Of particular importance to this paper is the work of McLaughlin et al. (\cite{McLaughlin2009}) which  is the first demonstration of reconnection {\emph{naturally}} driven by MHD wave propagation, via a process entitled {\emph{oscillatory reconnection}}.

MHD wave propagation in inhomogeneous media is a fundamental plasma process and the study of MHD waves in the neighbourhood of magnetic null points directly contributes to this area (see review by McLaughlin et al. \cite{McLaughlinREVIEW}). It is known that {\emph{null points}} - weaknesses in the magnetic field where the field strength, and hence the Alfv\'en speed, is zero - and {\emph{separatrices}} - topological features that separate regions of different magnetic flux connectivity - are an inevitable consequence of the distributed isolated magnetic flux sources at the photospheric surface, where the number of such null points will depend upon the magnetic complexity of the photospheric flux distribution (see, e.g., review by Longcope \cite{L2005}, and Close et al. \cite{Close2004}; R{\'e}gnier et al. \cite{RPH2008}; Longcope \& Parnell \cite{LP2009} for the statistics of coronal null points). It is also now known that MHD wave perturbations are omnipresent in the corona (e.g. Tomczyk et al. \cite{Tomczyk2007}). Thus, these two areas of scientific study; MHD waves and magnetic topology, {\emph{will}} encounter each other in the corona, i.e. MHD waves will propagate into the neighbourhood of coronal null points (e.g. blast waves from a flare will at some point encounter a null point).

McLaughlin \& Hood (\cite{MH2004}; \cite{MH2005}; \cite{MH2006a}; \cite{MH2006b}) investigated the behaviour of linear MHD waves (fast and slow magnetoacoustic waves and Alfv\'en waves) in the neighbourhood of a variety of 2D null points. It was found that the (linear) fast wave is focused towards the null point by a refraction effect and all the wave energy, and thus current density, accumulates close to the null, i.e. {\emph{null points will be locations for preferential heating by (linear) fast waves}}. The Alfv\'en wave propagates along magnetic fieldlines and so accumulates along the separatrices (in 2D) or along the spine or fan-plane (in 3D). Waves in the neighbourhood of a single 2D null point have also been investigated using cylindrical models, in which  the generated waves encircled the null point (e.g.  Bulanov \& Syrovatskii \cite{Bulanov1980};   Craig \& McClymont \cite{CraigMcClymont1991}; \cite{CraigMcClymont1993}; Craig \& Watson \cite{CraigWatson1992}; Hassam \cite{Hassam1992}) and it was found that the wave propagation leads to an  exponentially-large increase in the current density (see also Ofman \cite{Ofman1992}; Ofman et al. \cite{Ofman1993}; Steinolfson et al. \cite{Steinolfson1995} and a comprehensive review by  McLaughlin et al. \cite{McLaughlinREVIEW} for further details). 3D MHD wave activity about coronal null points has been investigated by various authors (e.g. Galsgaard et al. \cite{Galsgaard2003};   Pontin \& {Galsgaard} \cite{P1};  Pontin et al. \cite{P2}; McLaughlin et al. \cite{MFH2008}; Galsgaard \& Pontin \cite{klaus2011a}; \cite{klaus2011b}; Thurgood \& McLaughlin \cite{Thurgood2012}).

Reconnection can occur when strong currents cause the magnetic fieldlines to diffuse through the plasma and change their connectivity (Parker \cite{Parker}; Sweet \cite{Sweet}; Petschek \cite{Petschek}).  However, these papers did not include the effect of gas pressure, which would act to limit the growth of the current density. In considering the relaxation of a 2D X-type neutral point disturbed from equilibrium,   Craig \& McClymont (\cite{CraigMcClymont1991}) found that free magnetic energy is dissipated by a physical mechanism which couples resistive diffusion at the null to global advection of the outer field, which they called {\emph{oscillatory reconnection}}. An example of oscillatory reconnection generated by magnetic flux emerging into a coronal hole  was reported by Murray et al. (\cite{Murray2009}) who found a series of ``reconnection reversals'' take places as the system searches for equilibrium, i.e. the system demonstrates oscillatory reconnection in a self-consistent manner. The physics behind oscillatory reconnection has been investigated by McLaughlin et al. (\cite{McLaughlin2009}),   Murray et al. (\cite{Murray2009}) and Threlfall et al. (\cite{Threlfall2012}).

%Longcope \& Priest (\cite{LP2007}) investigated the diffusion of a 2D current sheet  subject to suddenly enhanced resistivity. They found that the diffusion couples to a fast MA mode which propagates the current  away at the local Alfv\'en speed.

%An example of oscillatory reconnection generated by flux emergence within a coronal hole was recently detailed by Murray et al. (\cite{Murray}). Finally, Longcope \& Priest (\cite{LP2007}) investigated the diffusion of a 2D current sheet  subject to suddenly enhanced resistivity. They found that the diffusion couples to a fast MA mode which propagates the current  away at the local Alfv\'en speed.

%Note that these transverse motions have been called Alfv\'en waves by some authors,  although this is subject to debate and they are  alternatively interpreted as kink waves (e.g. see arguments by  Erd\'elyi \& Fedun \cite{RF2007}; Van Doorsselaere et al. \cite{Van2008}). The dispute rests not with the observations themselves, but with the appropriate interpretation: MHD wave modes of an overdense cylinder versus MHD waves of a homogeneous plasma.

McLaughlin et al. (\cite{McLaughlin2012}) investigated the long-term evolution of an initially-buoyant magnetic flux tube emerging into a gravitationally-stratified coronal hole environment and reported on the resulting oscillations and outflows. They found that the physical mechanism of oscillatory reconnection {\emph{naturally}} generates quasi-periodic vertical outflows with a transverse/swaying aspect. There is currently a great deal of interest in observations of transverse motions in the solar atmosphere (e.g. Tomczyk et al. \cite{Tomczyk2007}; De Pontieu et al. \cite{Bart2007}; \cite{Bart2011}; Cirtain et al. \cite{Cirtain2007}; Erd\'elyi \& Taroyan \cite{Robertus2008}; Nishizuka et al. \cite{Nishizuka2008};  \cite{Nishizuka2011}; He et al. \cite{He2009a}; \cite{He2009b}; Liu et al. \cite{Liu2009}; \cite{Liu2011}; McIntosh et al. \cite{McIntosh2011};  Okamoto \& De Pontieu \cite{OBart2011}; Morton et al. \cite{Morton2012}; Yurchyshyn et al. \cite{Yurchyshyn2012}) and transverse/swaying motions have been observed over a range of wavelengths, speeds, temperatures and scales. However, the origin of these propagating, transverse oscillations remains a mystery, and these authors often note that the challenge remains to understand how and where these waves are generated in the solar atmosphere. Liu et al. (\cite{Liu2011}) summarises possible generation mechanisms for these transverse motions, including an oscillating wake from a coronal mass ejection or periodic reconnection (see, e.g., Chen \& Priest \cite{CP2006}; Sych et al. \cite{Sych2009}). The physical mechanism of oscillatory reconnection is another  possible source  of  these transverse motions. As reported by McLaughlin et al. (\cite{McLaughlin2012}), the transverse behaviour seen in the periodic jets  originating from the reconnection region of the inverted Y-shaped  structure is specifically due to the oscillatory reconnection mechanism, and would be absent for a single, steady-state reconnection jet. The physical mechanism also  naturally generates periodic outputs even though no periodic driver is imposed on the system.

Thus, there is a clear interest in furthering our understanding of the periodic nature of oscillatory reconnection. In this paper, we investigate the periodic signal generated by the mechanism, with a specific interest in measuring periods and decay rates as well as the robustness of our results, i.e. how do the results vary with the strength of the driver.

\subsection{Overview of McLaughlin et al. (2009)}\label{section1.1}

This paper will closely follow the work of McLaughlin et al. (\cite{McLaughlin2009}) as we investigate the periodic nature of oscillatory reconnection. These authors investigated the behaviour of nonlinear fast magnetoacoustic waves near  a 2D X-type neutral point and found that the incoming wave deforms the null point into a cusp-like point which in turn collapses to a current sheet. The system then evolves periodically through a series of horizontal/vertical current sheets with associated changes in connectivity, i.e. the system demonstrates the mechanism of oscillatory reconnection.

% Note that there is nothing special about the orientation of the first current sheet being horizontal followed by a vertical, this simply results from our particular choice of initial condition, and McLaughlin et al. (\cite{McLaughlin2012}) use the terminology {\emph{orientation 1}} and {\emph{orientation 2}}.

More specifically, McLaughlin et al. (\cite{McLaughlin2009})  found that the incoming (fast) wave propagates across the magnetic fieldlines and the initial profile, an annulus, contracts as the wave approaches the null point. This is the refraction behaviour that {{is}} typical of fast wave behaviour around magnetic null points (see, e.g., McLaughlin et al. \cite{McLaughlinREVIEW}) and results from the spatially-varying (equilibrium) Alfv\'en-speed profile.

The incoming wave was observed to develop discontinuities (for a physical explanation, see Appendix B of McLaughlin et al. \cite{McLaughlin2009} or, alternatively, Gruszecki et al. \cite{Marcin2011}) and these discontinuities form fast oblique magnetic shock waves, where the shock makes ${\bf{B}}$ refract away from the normal. Interestingly, the shock locally heats the initially $\beta=0$ plasma, creating $\beta \neq 0$ at these locations.

At a later time, the shocks overlap, forming a shock-cusp, which leads to the development of hot jets and in turn these jets substantially heat the local plasma and significantly deform the local magnetic field. By the time the shock waves reach the null, the (initially X-point) magnetic field has been deformed such that the separatrices  now touch one another rather than intersecting at a non-zero angle (Priest \& Cowley \cite{PC1975} call this \lq{cusp-like}\rq{}). The osculating field structure continues to collapse and forms a horizontal current sheet. However, the separatrices continue to evolve: the jets at the ends of the (horizontal) current sheet continue to heat the local plasma, which in turn expands. This expansion squashes and shortens the current sheet, forcing the separatrices apart. The (squashed) current sheet thus returns to a \lq{cusp-like}\rq{} null point that, due to the continuing expansion of the heated plasma, in turn forms a vertical current sheet. In effect, the (net) restoring force acts to return the  (deformed) null point  to its equilibrium state, but overshoots the equilibrium. The phenomenon then repeats itself: jets heat the plasma at the ends of this newly-formed (vertical) current sheet, the local plasma expands, the (vertical) current sheet is shortened, the system attempts to return itself to equilibrium, overshoots and forms a (second) horizontal current sheet. The evolution proceeds through a series of horizontal and vertical current sheets, and the system clearly displays oscillatory behaviour. It is also interesting to note that the final state is non-potential, where this is because the plasma to the left and right of the null  is (locally) hotter than that above and below. Consequently, a thermal-pressure gradient exists and causes the X-point to be slightly closed up in the vertical direction (i.e. generating a small, positive current). It is important to note that the non-potential final state is still in force balance and will eventually return to a potential state, but on a far greater timescale than our simulations ($t_{{\rm{diffusion}}} \sim  R_m = 10^4$ Alfv\'en times, where $R_m$ is the magnetic Reynolds number).  We also note that there is nothing unique about the orientation of the first current sheet being horizontal followed by a vertical, this simply results from the particular choice of initial condition, and McLaughlin et al. (\cite{McLaughlin2012}) use the more general terminology {\emph{orientation 1}} and {\emph{orientation 2}}.

%The hot jets set-up standing slow oblique magnetic shock waves arising from the shock-cusp

McLaughlin et al. (\cite{McLaughlin2009}) also present evidence of reconnection occurring in the system; reporting both a change in fieldline connectivity (qualitative evidence) and changes in the vector potential which directly showed a cyclic increase and decrease in magnetic flux on either side of the separatrices (see their Figures 12 and 13). Hence, since the system displayed both oscillatory behaviour {\emph{and}} reconnection, it was concluded that the system demonstrated the phenomenon of {{oscillatory reconnection}}.

% (see also Craig \& McClymont \cite{CraigMcClymont1991}; Murray et al. \cite{Murray2009}; McLaughlin et al. \cite{McLaughlin2012}; Threlfall et al. \cite{Threlfall2012}).

Our paper has the following outline: the basic setup, equations and assumptions are described in \S\ref{section2}, the periodic nature of oscillatory reconnection is detailed in \S\ref{section4} and the conclusions are given in \S\ref{section:conclusions}.

%This paper investigates the periodicity of the resultant signal. Specifically, we are interested in three questions:
%General interest in periodicity
%Thus, there is a growing interest in the generation mechanism of oscillatory reconnection.
%In this paper, we investigate the nature of the resultant periodic signal.

%Recently, great deal of interest in transverse oscillations and flows:
%Tomczyk et al (2007) – CoMP – ubiquitous, transverse disturbances propagating along magnetic field lines. Authors do not report on how these waves are generated.
%De Pontieu et al (2007) – Hinode/SOT – transverse waves in chromosphere with amplitudes 10-30 km/s periods 100-500 seconds (dynamic spicules/jet-like extrusions).
%De Pontieu et al (2011) – link between chromospheric spicules/jets and coronal spicules/jets – authors report there are currently no models for what drives and heats the observed jets.
%McIntosh et al (2011) – SDO – transition region observations of transverse oscillations, outwardly propagating, amplitudes 20km/s, period 100-500 seconds – authors not that the challenge remains to understand how and where these waves are generated in the solar atmosphere.
%De Pontieu & McIntosh (2010) – Hinode/EIS – quasi-periodic upflows (50-150 km/s line-of-sight)

%----------------------------------------------------------------
%----------------------------------------------------------------
%----------------------------------------------------------------
%----------------------------------------------------------------
%----------------------------------------------------------------
\section{Basic Equations}\label{section2}
%----------------------------------------------------------------
%----------------------------------------------------------------
%----------------------------------------------------------------
%----------------------------------------------------------------
%----------------------------------------------------------------

We consider the nonlinear, compressible, resistive MHD equations:
\begin{eqnarray}
 \rho \left[ {\partial {\bf{v}}\over \partial t} + \left( {\bf{v}}\cdot\nabla \right) {\bf{v}} \right] &=&- \nabla p + \left( {{\frac{1}{\mu}}}   \nabla \times {\bf{B}}  \right)\times {\bf{B}}  \; ,\nonumber \\
 {\partial {\bf{B}}\over \partial t}  &=& \nabla \times \left ({\bf{v}}\times {\bf{B}}\right ) + \eta \nabla ^2  {\bf{B}}\; ,\nonumber \\
 \rho \left[{\partial {\epsilon}\over \partial t}  + \left( {\bf{v}}\cdot\nabla \right) {\epsilon}\right] &=& - p \nabla \cdot {\bf{v}} + {{\frac{1}{\sigma}}} \left| {\bf{j}} \right| ^2 + Q_{\rm{shock}} \; \nonumber  ,\\
{\partial \rho\over \partial t} + \nabla \cdot \left (\rho {\bf{v}}\right ) &=& 0\; , \label{MHDequations}  
\end{eqnarray}

where $\rho$ is the mass density, ${\bf{v}}$ is the plasma velocity, ${\bf{B}}$ the magnetic induction (usually called the magnetic field), $p$ is the plasma pressure,  $ \mu = 4 \pi \times 10^{-7} \/\mathrm{Hm^{-1}}$  is the magnetic permeability,  $\sigma$ is the electrical conductivity,  $\eta=1/ {\mu \sigma} $ is the magnetic diffusivity, $\epsilon= {p / \rho \left( \gamma -1 \right)}$ is the specific internal energy density,   $\gamma={5 / 3}$ is the ratio of specific heats and ${\bf{j}} = {{\nabla \times {\bf{B}}} / \mu}$ is the electric current density. 

We solve these governing equations numerically using a Lagrangian remap, shock-capturing code called {\emph{LARE2D}} (Arber et al. \cite{Arber2001}),   which utilizes artificial shock viscosity to introduce dissipation at steep gradients. The details of this technique, often called Wilkins  viscosity, can be found in Wilkins (\cite{Wilkins1980}). Thus, $Q_{\rm{shock}}$ represents the viscous heating at shocks. 

%Heat conduction and radiative effects are neglected in the present study.

% are solved numerically using a Lagrangian remap, shock-capturing code called {\emph{LARE2D}} (Arber et al. \cite{Arber2001}). 

We now introduce a change of scale to non-dimensionalise all variables. Letting ${\rm{\bf{v}}} = {\rm{v}}_0 {\mathbf{v}}^*$,  ${\mathbf{B}} = B {\mathbf{B}}^*$, $x = L x^*$, $y=L y^*$, $z = L z^*$, $\rho={\rho}_0 \rho^*$, $p = p_0 p^*$, ${\bf{j}}=j_0\:{\bf{j}}^*$, $\nabla = \nabla^* / L$, $t={t}_0 t^*$  and $\eta = \eta_0$, where * denotes a dimensionless quantity and ${\rm{v}}_0$, $B$, $L$, ${\rho}_0$, $p_0$, $j_0$,  ${t}_0$ and $\eta_0$ are constants with the dimensions of the variable they are scaling.  We then set $ {B} / {\sqrt{\mu \rho _0 } } ={\rm{v}}_0$ and ${\rm{v}}_0 =  {L} / {{t_0}}$ (this sets ${\rm{v}}_0$ as a constant background Alfv\'{e}n speed). We also set $j_0 = B/\mu L$ and  ${\eta_0 {t}_0 } /  {L^2} =R_m^{-1}$, where $R_m$ is the magnetic Reynolds number, and choose $R_m=10^4$. This process non-dimensionalises equations (\ref{MHDequations}) and under these scalings, $t^*=1$ (for example) refers to $t={t}_0=  {L} / {{\rm{v}}_0}$; i.e. the time taken to travel a distance $L$ at the background Alfv\'en speed.

%For the rest of this paper, we drop the star indices; the fact that all variables are now non-dimensionalised is understood.

%----------------------------------------------------------------
%----------------------------------------------------------------
%----------------------------------------------------------------
%----------------------------------------------------------------
%----------------------------------------------------------------

There is no fixed dimensional length scale to our X-point system (X-points are scale-free), and thus we have a great deal of freedom in choosing our dimensional constants. We choose  $L=1\:$Mm and  $B=1\:$G (for simplicity, and where these choices allow an intuitive understanding of equation \ref{Xpoint} below) and we chose a coronal density of ${\rho}_0= 5 \times 10^{-13}\:$kg/m$^3$ and coronal temperature of $T_0=10^6\:$K. This sets ${\rm{v}}_0={B}/\sqrt{\mu \rho_0} =  126.2\:$km/s, $t_0 = {L} / {{\rm{v}}_0} = 7.93 \:$seconds and $j_0 = B/\mu L = 8 \times 10^{-5}\:$A. The values returned from equations (\ref{MHDequations}) are made dimensional using these solar constants.

%we follow the choices of McLaughlin et al. (\cite{McLaughlin2012}) and  choose the following values for these dimensional constants:  (photospheric) pressure scale height $L=1.7\times 10^5\:$m, time $t_0=25\:$seconds, velocity ${\rm{v}}_0={L} / {{t_0}}=6.8\times 10^3\:$m/s, density ${\rho}_0=3.0\times 10^{-4}\:$kg/m$^3$, pressure  $p_0=1.2\times 10^4\:$Pa, temperature $T_0=5.6\times 10^3\:$K and magnetic field $B=1.3\times 10^3\:$G.
%The values returned from  equations (\ref{MHDequations}) are made dimensional using the following choice of solar constants:  (photospheric) pressure scale height $L=1.7\times 10^5\:$m, time $t_0=25\:$seconds, velocity ${\rm{v}}_0={L} / {{t_0}}=6.8\times 10^3\:$m/s, density ${\rho}_0=3.0\times 10^{-4}\:$kg/m$^3$, pressure  $p_0=1.2\times 10^4\:$Pa, temperature $T_0=5.6\times 10^3\:$K, magnetic field $B=1.3\times 10^3\:$G and gravity $g_0=270\:$m/s.

%----------------------------------------------------------------
%----------------------------------------------------------------
%----------------------------------------------------------------
%----------------------------------------------------------------
%----------------------------------------------------------------
\subsection{Basic equilibrium  and numerical set-up}\label{section:2.1}
%----------------------------------------------------------------
%----------------------------------------------------------------
%----------------------------------------------------------------
%----------------------------------------------------------------
%----------------------------------------------------------------

\begin{figure*}[t]
\begin{center}
\includegraphics[width=6.5in]{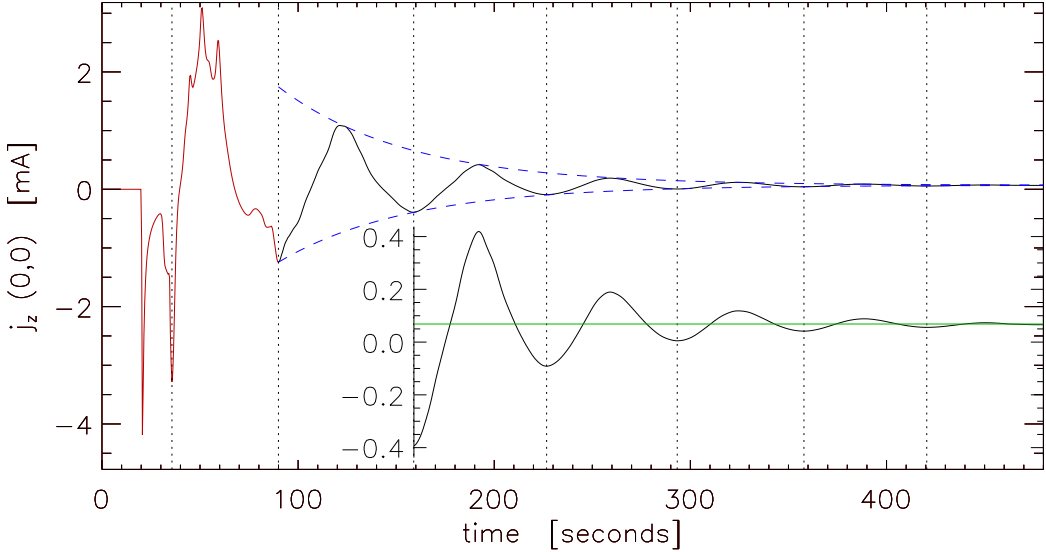}
\caption{Plot of time evolution of $j_z(0,0,t)$ (measured in milliAmps) for $0\le t\le 480\:$seconds. Red line indicates {\emph{impulsive phase}} and black line indicates {\emph{stationary}} phase. Insert shows the same time evolution over $158.9\le t\le 480\:$seconds (same horizontal axis, different vertical axis).  The black dotted lines indicate the formation times of all the horizontal current sheets and the green line indicates $j_{{\rm{final}}} = 0.8165j_0= 6.5\times 10^{-5}\:$A. The blue dashed lines indicate an exponentially-damped envelope ${\rm{max}}(j_z)|_{\rm{stationary}}\times e^{-\lambda t} +  j_{{\rm{final}}}$ and ${\rm{min}}(j_z)|_{\rm{stationary}}\times e^{-\lambda t} +  j_{{\rm{final}}}$  where $\lambda = -0.015\:{\rm{s}}^{-1}$.}
\label{figure1}
\end{center}
\end{figure*}

To set-up our system, we follow the numerical framework of McLaughlin et al. (\cite{McLaughlin2009}). Thus, we consider a simple 2D X-type neutral point as our equilibrium magnetic field, where the initial field is taken as:
\begin{eqnarray}
{\bf{B}}_0 = \frac{B}{L} \left(y, x, 0\right) \;,   \label{Xpoint}
\end{eqnarray}
where  $B=1\:$G is a characteristic field strength and $L=1\:$Mm is the length scale for magnetic field variations.  This magnetic field can be seen in McLaughlin et al. (\cite{McLaughlin2009}, their Figure 1).

%We take the equilibrium density to be uniform, i.e. $\rho=\rho_0$;  a spatial variation in $\rho_0$ can cause phase mixing (Heyvaerts \& Priest \cite{Heyvaerts1983}; De Moortel {et al.} \cite{DeMoortel1999}; Hood et al. \cite{Hood2002}). We set $R_m=10^4$.

Initially, we consider the equilibrium plasma to be cold: $T=0\:$K  (i.e. $\beta (t=0)=0$) and, hence, ignore plasma pressure effects. However, McLaughlin et al. (\cite{McLaughlin2009}) showed that magnetic shocks will heat the plasma  and so the plasma will not remain cold (see e.g. $\S 1.5$ in Priest \& Forbes \cite{magneticreconnection2000}).

In order to excite a pure  fast magnetoacoustic wave, we consider an initial condition that perturbs velocity {\emph{purely across}} the equilibrium magnetic field. Using the terminology of  McLaughlin et al. (\cite{McLaughlin2009}), there are three distinguishing  velocity components that can be considered in our system:
\begin{itemize}
\item[$\bullet$]{${\rm{v}}_\perp= {\bf{v}}\times{\bf{B}}_0 \cdot {\hat{\bf{z}}}={\rm{v}}_x B_y - {\rm{v}}_y B_x$, which corresponds to the velocity across the equilibrium field, and hence corresponds to the fast magnetoacoustic wave (the only MHD wave that can cross fieldlines)},
\item[$\bullet$]{ ${\rm{v}}_\parallel=  {\bf{v}}\cdot{\bf{B}}_0={\rm{v}}_x B_x + {\rm{v}}_y B_y$, which corresponds to the velocity along the equilibrium field and corresponds to the propagation of the slow magnetoacoustic wave},
\item[$\bullet$]{${\rm{v}}_z={\bf{v}}\cdot{\hat{\bf{z}}}$, which corresponds to the velocity in the invariant direction and hence corresponds to the Alfv\'en wave.}
\end{itemize}
Interestingly, these three velocity components - each isolating an individual MHD mode -  are in good agreement with those reported by Thurgood \& McLaughlin (\cite{Thurgood2012}) who used the equilibrium magnetic field and the flux function  (which is parallel to the invariant direction) to define an orthogonal coordinate system to isolate and identify the propagation of  each of the MHD modes. Using their convention, ${\rm{v}}_\perp = {\bf{v}} \cdot { {\bf{B}}}_0 \times {\hat{{\bf{A}}}}$, where  ${\bf{A}} = \frac{1}{2} \left(y^2-x^2\right) {\hat{\bf{z}}}$ is the flux function, and where  perturbations in the ${\bf{B}}_0  \times  \hat{ {\bf{A}}}$-direction were shown to correspond to those of the fast wave (for helicity-free systems, such as our 2D X-point).

%The utilisation of these three velocity variables aids in MHD mode interpretation and isolation and, as seen in McLaughlin \& Hood (\cite{MH2004}) led to an an analytical solution to the linear, cold plasma equations for ${\rm{v}}_\perp$.

%and  ${\rm{v}}_\parallel=  {\bf{v}}\cdot{\bf{B}}={\rm{v}}_x B_x + {\rm{v}}_y B_y$. Here, ${\rm{v}}_\perp$ and ${\rm{v}}_\parallel$ are related to the perpendicular and parallel velocity, respectively and, as seen in \cite{MH2004}, their implementation naturally simplifies the governing equations, aids in MHD mode interpretation  and  (for ${\rm{v}}_\perp$) led to an analytical solution to the linear, cold plasma equations.

To perturb our system, we consider an initial condition in velocity such that:
\begin{eqnarray}
{\rm{v}}_\perp \left(x,y,t=0 \right)  &=& 2C \sin \left[ \pi \left(r - 4.5 \right) \right]{\rm{\;\; for\;\;\;}}4.5\le r \le 5.5 \nonumber\;,\\
{\rm{v}}_\parallel \left(x,y,t=0 \right)  &=&{\rm{v}}_z \left(x,y,t=0 \right)= 0\;\label{ICs}
\end{eqnarray}
where $r^2=x^2+y^2$ and $2C$ is our initial amplitude and initial condition (\ref{ICs}) describes a circular, sinusoidal pulse in ${\rm{v}}_\perp$. Thus, as argued above, we initially generate (only) a pure fast wave in our system. {{Note that the  velocity profile prescribed by equation  (\ref{ICs}) appears as the symmetric  $m = 0$ mode in ${{\rm{v}}_\perp}$, but corresponds to the asymmetric $m=2$ mode in Cartesian components. This is why the first current sheet has horizontal orientation, as per \S\ref{section1.1}.}}

%where $r^2=x^2+y^2$ and $2C$ is our initial amplitude and initial condition (\ref{ICs}) describes a circular, sinusoidal pulse in ${\rm{v}}_\perp$. Thus, as argued above, we initially generate (only) a pure fast wave in our system. {\bf{Note that the  velocity profile prescribed by equation (\ref{ICs}) appears as the $m = 0$ mode in ${{\rm{v}}_\perp}$ but corresponds to the $m = 2$ mode in Cartesian components.}}

This initial pulse  will naturally split into two waves -  an outgoing wave and an incoming wave - each of amplitude $C$. In this paper, we will focus on the incoming wave, i.e. the wave propagating towards the null point. In this paper, we will conduct a parameter study of initial wave amplitude $C$. Note that setting $C=1$ recovers the results of McLaughlin et al. (\cite{McLaughlin2009}) and choosing a small value for $C$, say $C=0.001$, recovers the the linear results from McLaughlin \& Hood ({\cite{MH2004}}; i.e. see Appendix A of McLaughlin et al. \cite{McLaughlin2009}). Under our dimensionalisation, a choice of $C=1$ corresponds to an incoming  wave with maximum initial {{Cartesian}} velocity  ${{\rm{v}}_\perp}/r = {\rm{v}}_0/5 =  25.2 \:$km/s at a distance of $r=5L= 5\:$Mm and an equilibrium  magnetic field strength of $5\:$G.

%In cartesian coordinates: ${\rm{v}}_x = \left( {\rm{v}}_\parallel B_x + {\rm{v}}_\perp B_y \right) / \left| {\bf{B}}\right|^2$ and ${\rm{v}}_y = \left( {\rm{v}}_\parallel B_y - {\rm{v}}_\perp B_x \right) / \left| {\bf{B}}\right|^2$. In polar coordinates: ${\rm{v}}_x = {\rm{v}}_\perp \cos{\theta} / r$, ${\rm{v}}_y = -{\rm{v}}_\perp \sin{\theta} / r$, where $r=\sqrt{x^2+y^2}$ and we take ${\rm{v}}_\parallel$ to be initially zero. We note that with our choice of magnetic null point (equation \ref{Xpoint}), if we drive any of the velocity variables, the system will naturally develop a $\theta$ dependence.

The governing equations (\ref{MHDequations}) with initial conditions (\ref{ICs}) are  solved computationally in a square domain $x,y   \in [-20,20]$ with a numerical resolution of $6144 \times 6144$.  Zero gradient boundary conditions are applied to the variables ${\bf{B}}$, $\rho$, $\epsilon$ at the four boundaries, and  ${\bf{v}}$ is set to zero on all boundaries, i.e. reflective boundaries. A numerical damping region exists for $x^2+y^2 \ge 6$ {{which gradually removes kinetic energy from the outgoing waves}} and so all oscillations that enter this region are slowly damped away, and {{hence they do not influence the behaviour about the null}}. The  (equilibrium) Alfv\'en speed increases with distance from the null point and, hence, waves accelerate as they propagate outwards. Since we do not want reflected waves to influence our null point, implementation of such a damping region is essential.

%umerical damping region exists for x2 + y2 ≥  6 *** which gradually removes kinetic energy from the outgoing waves***  and so all oscillations that enter this region are slowly damped away, ***and so they do not influence the behaviour at the null ***. The (equilibrium) Alfv´en speed increases with distance from the null point and, hence, waves accelerate as they propagate outwards. Since we do not want reflected waves to influence our null point, implementation of such a damping region is essential.

%%%%%%%%%%%%%%%%%%%%%%%%%%%%%%%%%%%%%%%%%%%%%%%%%%%%%%%%%%%%%%%%%%%%%%%%%%%%%%%%%%%%%%

%----------------------------------------------------------------
%----------------------------------------------------------------
%----------------------------------------------------------------
%----------------------------------------------------------------
%----------------------------------------------------------------
\section{Aperiodic driver leading to periodic behaviour}\label{section4}
%----------------------------------------------------------------
%----------------------------------------------------------------
%----------------------------------------------------------------
%----------------------------------------------------------------
%----------------------------------------------------------------

% We find that the incoming (fast) wave propagates across the magnetic fieldlines and the initial profile, an annulus, contracts as the wave approaches the null point. This is the refraction behaviour that is typical of fast wave behaviour around magnetic null points (see, e.g., McLaughlin et al. \cite{McLaughlinREVIEW}) and results from the spatially-varying, equilibrium Alfv\'en-speed profile.

We set $C=1$ in equations (\ref{ICs}) and, as expected, we recover the results of McLaughlin et al. (\cite{McLaughlin2009}) and {{readers are directed to that paper for full details (primarily their Figures 2, 5 \& 6) and also this paper's $\S\ref{section1.1}$.}} One of the key results from McLaughlin et al. (\cite{McLaughlin2009}) was the production of a {\emph{periodic}} response resulting from an {\emph{aperiodic}} input, i.e. the physical mechanism of oscillatory reconnection naturally gave rise to periodic behaviour. This periodic response can be quantitatively measured by analysing the time evolution of the (electric) current density, specifically ${\bf{j}}= (0,0,j_z)$, at the null point itself. This can be seen in  McLaughlin et al. (\cite{McLaughlin2009}, their Figure 10) and the analysis of this time signal is the primary focus of this paper. In recreating the simulations of  McLaughlin et al. (\cite{McLaughlin2009}), we recover this periodic  time series, but at a higher numerical resolution, $6144\times6144$, and at a higher cadence, $dt=0.05\:t_0$. The time evolution of  $j_z(0,0,t)$ for $0\le t \le 480\:$s can be seen in Figure \ref{figure1}. Note that due to the symmetry of our system, the null is {\emph{always}} located at the origin.

%It is clear that there is oscillatory behaviour during  the evolution of $j_z(0,0,t)$ 

We identify two distinct regimes in Figure \ref{figure1}:  $0\le t < 90\:$s (red line in Figure \ref{figure1}) which we refer to as the {\emph{impulsive phase}} and $t \ge 90\:$s (black line in Figure \ref{figure1}) which we refer to as the {\emph{stationary phase}}. The impulsive (or transient) phase is a spiky, irregular  signal, with multiple local extrema indicating: the arrival time of the shock-cusp at the null point (at $t=20.6\:$s), the formation of the first horizontal current sheet (at $t=35.7\:$s, $j_z<0$), the formation of the first vertical current sheet (at $t=59.1\:$s, $j_z>0$) and the formation of the second horizontal current sheet (at  $t=90.0\:$s). The (current-sheet) signal is further contaminated  due to addition of small currents related to the propagation of shock waves across the null point. We define the end of the impulsive phase as the formation time of the second horizontal current sheet, which in our $C=1$ simulation occurs at $t=90.0\:$s.

%Identification of these extrema is only possible by comparison with the evolution of the separatrices and contours of ${\bf{v}}$.

After $t=90.0\:$s, the evolution of $j_z(0,0,t)$ is much cleaner and closer to (damped) sinusoidal. For $t \ge 90.0\:$s, the extrema exactly match the formation of the cyclic current sheets, with $j_z<0\:$/$\:>0$ indicating horizontal/vertical current sheets, respectively. We define this \lq{cleaner signal}\rq{ } regime as the {\emph{stationary phase}}, i.e. the regime characterised by a (damped) sinusoidal signal after the transients of the impulsive phase have leaked away,  and this phase starts at the formation time of the second horizontal current sheet. The black dotted lines in Figure \ref{figure1} indicate the formation times of all the horizontal current sheets (for both phases).

Note that in labelling these two regimes, i.e  the {{impulsive phase}} and the   {{stationary phase}}, we have adopted the terminology usually associated with the excitation and damping of trapped and leaky modes in coronal loop oscillations (see, e.g., Terradas et al. \cite{Terradas2005}; \cite{Terradas2006}; Luna et al. \cite{Luna2008}; McLaughlin \& Ofman \cite{MO2008} and reference therein).

\begin{figure*}[t]
\begin{center}
\includegraphics[width=6.5in]{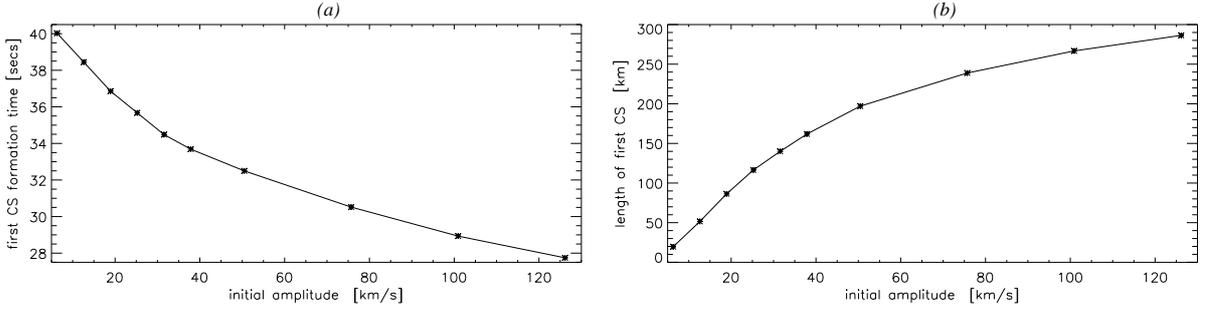}
\caption{Plot of $(a)$ formation time (measured in seconds) and $(b)$ length (measured in km and evaluated at the formation time) of first horizontal current sheet  versus initial velocity amplitude (measured in km/s), where ${\rm{v}}_0=25.2\:$km/s corresponds to $C=1$ simulation.}
\label{figure2}
\end{center}
\end{figure*}

%----------------------------------------------------------------
%----------------------------------------------------------------
%----------------------------------------------------------------
%----------------------------------------------------------------
%----------------------------------------------------------------
\subsection{Impulsive Phase}\label{section4.1}
%----------------------------------------------------------------
%----------------------------------------------------------------
%----------------------------------------------------------------
%----------------------------------------------------------------
%----------------------------------------------------------------

Let us consider the impulsive phase of the evolution of $j_z(0,0,t)$, i.e. evolution over $0\le t < 90$ (indicated by  red line in Figure \ref{figure1}). We see that $j_z(0,0,t)$ is zero (since the equilibrium is potential) until the arrival of the fast oblique magnetic shock (which has its own associated current density) at the null point,  indicated by the first (negative) extrema at $t=20.6\:$s. This is followed by a second (local) minimum at $t=35.7\:$s, which  is the formation time of the first horizontal current sheet (i.e. not at $t=20.6\:$s). The correspondence between extrema and current-sheet formation cannot be deduced from Figure \ref{figure1} alone, but can be determined by comparison with the evolution of the separatrices and contours of ${\bf{v}}$. The value of $j_z(0,0,t=35.7)=-41.25 j_0=-3.3\:$mA is proportional to the length of the first horizontal current sheet, which  is measured directly from the numerical simulation as $0.1162\: L = 116.2\:$km (for the $C=1$ simulation).

The formation time and length (at formation time) of the first horizontal current sheet is of key interest here, since this directly reflects the driver (akin to a forcing term) of the system. We investigate how the formation time and length of the first horizontal current sheet vary as functions of the initial driving amplitude (from equation \ref{ICs}). Figure \ref{figure2}a shows the  formation time (measured in seconds) of the first horizontal current sheet as a function of initial velocity amplitude (measured in km/s), where ${\rm{v}}_0=25.2\:$ corresponds to $C=1$ simulation. We see that the {\emph{greater}} the driving amplitude, the {\emph{earlier}} the first current sheet forms. This is intuitive as we expect  fast oblique magnetic shocks with a larger amplitude to propagate faster and thus reach the null point more rapidly, compared to waves driven with a smaller amplitude.

Figure \ref{figure2}b shows how the current sheet length (measured in km) evaluated at the formation time of the first horizontal current sheet  varies as a function of initial velocity amplitude. We see that the {\emph{greater}} the driving amplitude, the  {\emph{longer}} the  horizontal current sheet. The length of the current sheet is also directly proportional to the value of $j_z(0,0)$, and so we also conclude that the {\emph{greater}} the driving amplitude, the {\emph{stronger}} the value of  $|j_z(0,0)|$ at the corresponding time. Again, this result is intuitive; it is the fast oblique magnetic shock that physically deforms the   X-point into an osculating field structure (and ultimately into a  horizontal current sheet) and thus we would expect {\emph{stronger}} fast oblique magnetic shocks (i.e. with a larger amplitude) to deform, i.e. refract ${\bf{B}}$ away from the normal, and \lq{squash}\rq{} the magnetic field to a greater extent, and thus to form {\emph{longer}} and {\emph{stronger}} current sheets.

%for the length of the current sheet, we conclude that the {\emph{greater}} the driving amplitude, the  {\emph{longer}} the 
%We see that the {\emph{greater}} the driving amplitude, the {\emph{stronger}} the value of  $|j_z(0,0)|$ and, since the value of $j_z(0,0)$ acts a proxy for the length of the current sheet, we conclude that the {\emph{greater}} the driving amplitude, the  {\emph{longer}} the current sheet 
%Note that in our system, 
%$j_z(0,0)<0$ for horizontal current sheets and so we have plotted absolute magnitude for clarity
%how the absolute magnitude of $j_z(0,0)$ evaluated at the formation time of the first (horizontal) current sheet

\begin{figure}
\begin{center}
\includegraphics[width=3.75in]{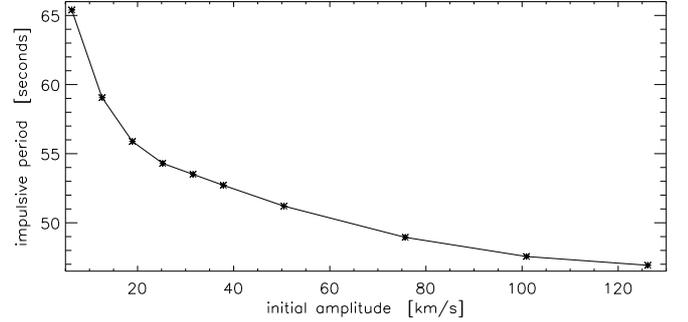}
\caption{Plot of impulsive period, i.e. time taken to evolve from the first horizontal current sheet to the second horizontal current sheet (measured in seconds) versus initial velocity amplitude (measured in km/s), where ${\rm{v}}_0=25.2\:$km/s corresponds to $C=1$ simulation.}
\label{figure4}
\end{center}
\end{figure}

\begin{figure*}[t]
\begin{center}
\includegraphics[width=6.5in]{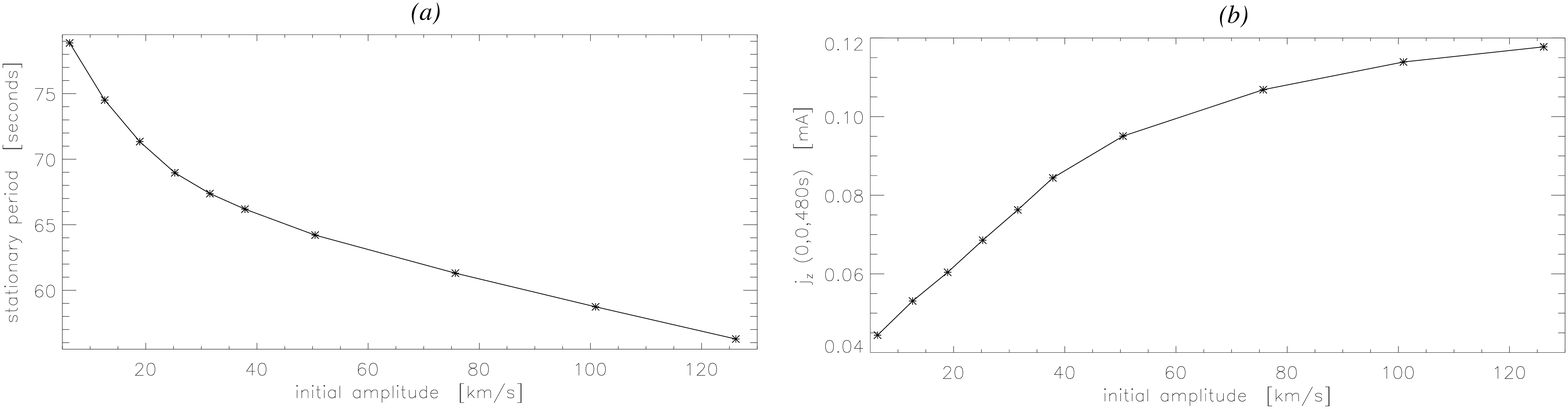}
\caption{Plot of $(a)$ stationary period, i.e. the time taken  to evolve from the second  horizontal current sheet to the third (measured in seconds) and $(b)$  $j_{{\rm{final}}}$, i.e. $j_z(0,0)$ evaluated at $t=480\:$s, (measured in milliAmps)  versus initial velocity amplitude (measured in km/s), where ${\rm{v}}_0=25.2\:$km/s corresponds to $C=1$ simulation.}
\label{figure5}
\end{center}
\end{figure*}

Finally, let us investigate the time taken to evolve from the first horizontal current sheet (at $t=35.7\:$s) to the formation time of the second horizontal current sheet (at $t=90.0\:$s for the $C=1$ simulation), namely the time taken for one complete cycle (i.e. horizontal current sheet evolves to vertical, evolves back to horizontal). This time, which we refer to as the {\emph{impulsive period}}, is calculated to be  $t=54.3\:$s (for initial amplitude ${\rm{v}}_0=25.2\:$km/s in the $C=1$ system). Again, we now investigate how this impulsive period varies with the initial driving amplitude ($\propto C {\rm{v}}_0$) and this can be seen in Figure \ref{figure4}. Here, we see that the {\emph{greater}} the initial driving amplitude, the {\emph{shorter}} the resulting impulsive period, i.e. the shorter the time taken  to evolve from the first horizontal current sheet to the second. Recalling the dependency seen Figure \ref{figure2}b, this means that for {\emph{longer}} current sheets, the impulsive period is {\emph{shorter}}. This means that the restoring force must be stronger and so we conclude that {\emph{longer}} current sheets have a {\emph{stronger}} restoring force. In this way, the system acts as a {\emph{harmonic oscillator}}, i.e. the greater the displacement away from equilibrium, the stronger the restoring force.

%But... aperiodic motions lead to periodic motions? How?   What is the generation mechanism?
%We find that the physical mechanism of oscillatory reconnection naturally generates quasi-periodic vertical outflows, with a transverse/swaying aspect.
%Self- consistent => no periodic driver but physical mechanism gives rise to periodic behaviour.

%----------------------------------------------------------------
%----------------------------------------------------------------
%----------------------------------------------------------------
%----------------------------------------------------------------
%----------------------------------------------------------------
\subsection{Stationary Phase}\label{section4.2}
%----------------------------------------------------------------
%----------------------------------------------------------------
%----------------------------------------------------------------
%----------------------------------------------------------------
%----------------------------------------------------------------

%\caption{Plot of stationary period, i.e. the time taken  to evolve from the second  horizontal current sheet to the third (measured in seconds)  versus initial velocity amplitude (measured in km/s), where ${\rm{v}}_0=25.2\:$km/s corresponds to $C=1$ simulation.}
%\begin{figure}
%\begin{center}
%\includegraphics[width=3.75in]{XXXXFIG5.eps}
%\caption{Plot of $j_{{\rm{final}}}$, i.e. $j_z(0,0)$ evaluated at $t=480\:$s, (measured in $10^{-3}\:$Amps) versus initial velocity amplitude (measured in km/s), where ${\rm{v}}_0=25.2\:$km/s corresponds to $C=1$ simulation.}
%\label{figure6}
%\end{center}
%\end{figure}

Let us now investigate the  {\emph{stationary phase}} of the oscillation seen in Figure \ref{figure1} (black line), i.e. the time evolution of $j_z(0,0,t)$ for $t_2 \le t\le 480\:$s, where $t_2$ is the formation time of the second horizontal current sheet ($t_2=90.0\:$s for the $C=1$ simulation).    Figure \ref{figure1} also  has an insert showing the (same)  time evolution of $j_z(0,0,t)$ for $158.9\le t\le 480\:$s, i.e. same horizontal time axis, different vertical axis. Note that the insert does not show the start of the stationary phase, only a later part of it (starting at the time of the third horizontal current sheet). The green line indicates $j_{{\rm{final}}}$, i.e. the finite amount of  current density left in the system at $t=480\:$s when the system has reached its final, non-potential state. For the $C=1$ system, this final state occurs at $t=480\:$s (8 mins) and  $j_{{\rm{final}}} = 0.8165j_0=6.5\times 10^{-5}\:$A.

We see that there is clear oscillatory behaviour in the stationary phase and, moreover, the oscillation is {\emph{exponentially decaying}}. This can be seen in Figure \ref{figure1} and the blue dashed lines indicate an exponentially-damped envelope  ${\rm{max}}(j_z)|_{\rm{stationary}}\times e^{-\lambda t} +  j_{{\rm{final}}}$ and ${\rm{min}}(j_z)|_{\rm{stationary}}\times e^{-\lambda t} +  j_{{\rm{final}}}$  where $\lambda = -0.015\:{\rm{s}}^{-1}$  is determined experimentally.

%This exponential damping is also observed for all values of $C$.

% In the stationary phase, the evolution of $j_z(0,0,t)$ is much cleaner and closer to sinusoidal than in the impulsive/transient phase. Moreover, in the stationary phase  the extrema exactly match the formation of the cyclic current sheets, with $j_z<0$/$j_z>0$ indicating horizontal/vertical current sheets, respectively, and the black dotted lines in Figure \ref{figure1} indicate the formation times of all the horizontal current sheets (for both phases). 

Let us now investigate the period associated with the stationary phase, which we define as the time taken to evolve between horizontal current sheets. Specifically, we define the {\emph{stationary period}} as the time taken to evolve from the second horizontal current sheet to the third, i.e. the first complete oscillation within the stationary phase.  For the $C=1$ simulation, these formation times are $90.0\:$s and $158.9\:$s resulting in a stationary period of $69.0\:$s (note we present results here correct to 1 decimal places, but calculate periods to a greater degree of accuracy). Similar results are obtained for alternative definitions of the stationary periods, e.g. time taken to evolve from one vertical current sheet to the next.

% (note that the  second horizontal current sheet of all phases in the first of the stationary phase,
% (which is actually  the first current sheet of the stationary phase) to the thrid horizontal current sheet
% (the second horizontal current sheet of the stationary phase, which is actually the third horizontal current sheet when phases are ignored).

We now investigate how the stationary period varies  with the initial driving amplitude and this can be seen in Figure \ref{figure5}a.  Here, we see that the {\emph{greater}} the initial driving amplitude, the {\emph{shorter}} the resulting stationary period. Thus, these results are in agreement with those in $\S \ref{section4.1}$, i.e. the  {\emph{stronger}} the initial driving amplitude, the {\emph{longer}} the resulting current sheet, thus the {\emph{stronger}} the restoring force, thus the {\emph{shorter}} the resulting period. Coupled with the exponential decay, we see that in the stationary phase, the system acts akin to a  {\emph{damped harmonic oscillator}}.

%, i.e. the greater the displacement away from equilibrium, the stronger the restoring force.
%haven't presented figure showning this is trully valid in stationary phase, i.e. length of second CS versus time.

%It is important to note that there is a quantitative  difference between the impulsive period ($t=54.3\:$s for $C=1$) and the stationary period ($t=69.0\:$s for $C=1$), and that the stationary period is longer than the corresponding impulsive period for all investigated values of $C$. 

% This indicates that different physical processes are dominant in each phase, and thus validates our approach of dividing the evolution into two distinct phases.

We also measure all the proceeding  oscillations in the stationary phase, i.e. time taken to evolve from the second/third horizontal current sheet to the third/fourth horizontal current sheet, and obtain similar periods of  $67.8\:$s and $66.6\:$s respectively. Interestingly, this means that the stationary period appears to be slightly decreasing by roughly $1.8\%$ per oscillation.

%{Plot of $j_{{\rm{final}}}$, i.e. $j_z(0,0)$ evaluated at $t=480\:$s, (measured in $10^{-3}\:$Amps) versus initial velocity amplitude (measured in km/s), where ${\rm{v}}_0=25.2\:$km/s corresponds to $C=1$ simulation.}
% $j_{{\rm{final}}} = 0.8165j_0=6.5\times 10^{-5}\:$A for the $C=1$ system.

Finally, we investigate $j_{{\rm{final}}}$, the finite amount of  current density left in the system  when the system has reached its final, non-potential state and this can be seen in  Figure \ref{figure5}b. For the $C=1$ system, $j_{{\rm{final}}} = 0.8165j_0=6.5\times 10^{-5}\:$A which is measured at $t=480\:$s. In all our simulations, $j_{{\rm{final}}}>0$ indicating that the (final) X-point is very slightly closed up in the vertical direction, i.e. $j_z>0$ is associated with vertical current sheets. This is because the (local) plasma to the left and right of the X-point is slightly hotter, since that is where the initial, strongest jet heating occurs. Thus, the existence of this  thermal-pressure gradient coupled with force balance requires the final state to be non-potential.

From Figure \ref{figure5}b, we see that the {\emph{stronger}} the initial driving amplitude, the {\emph{greater}} the value of $j_{{\rm{final}}}$. Again, this is intuitive: fast oblique magnetic shocks with a {\emph{larger}} amplitude will  intersect  to form {\emph{stronger}}, {\emph{hotter}} jets to the left and right of the  X-point. Thus, this local plasma will be hotter at the end of the simulation, indicating a stronger thermal-pressure gradient and thus, in order to achieve force balance, a greater absolutely value of the Lorentz force, i.e. a {\emph{greater}} value of $j_{{\rm{final}}}$.

%%%%%%%%%%%%%%%%%%%%%%%%%%%%%%%%%%%%%%%%%%%%%%%%%%%%%%%%%%%%%%%%%%%%%%%%%%%%%%%%%%%%%%
%%%%%%%%%%%%%%%%%%%%%%%%%%%%%%%%%%%%%%%%%%%%%%%%%%%%%%%%%%%%%%%%%%%%%%%%%%%%%%%%%%%%%%

%----------------------------------------------------------------
%----------------------------------------------------------------
%----------------------------------------------------------------
%----------------------------------------------------------------
%----------------------------------------------------------------
\section{Conclusions}\label{section:conclusions}
%----------------------------------------------------------------
%----------------------------------------------------------------
%----------------------------------------------------------------
%----------------------------------------------------------------
%----------------------------------------------------------------

%This paper describes an investigation into the nature of nonlinear fast magnetoacoustic waves in the neighbourhood of a 2D magnetic X-point. We have solved the compressible and resistive MHD equations using a Lagrangian remap, shock capturing code ({\emph{LARE2D}}). We consider a circular, sinusoidal pulse in ${\rm{v}}_\perp$ as our initial condition in velocity (equation \ref{ICs}), which naturally splits into two waves and we focus on the wave travelling towards the null point. Implemented damping regions remove the kinetic energy from the outgoing wave. Initially, we consider an incoming wave of amplitude $C=1$.

This paper describes an investigation into the periodicity of oscillatory reconnection, specifically oscillatory reconnection initiated by a nonlinear fast magnetoacoustic wave deforming a 2D magnetic X-point. We have solved the compressible, resistive, nonlinear  MHD equations using a Lagrangian remap, shock-capturing code ({\emph{LARE2D}}) and have followed the numerical set-up of McLaughlin et al. (\cite{McLaughlin2009}). As in that paper, we find that the fast magnetoacoustic wave develops into a fast oblique magnetic shock wave which significantly deforms the local magnetic fieldlines, to the extent that  the incoming wave deforms the null point into a cusp-like point which in turn collapses to a current sheet. The system then evolves periodically through a series of horizontal and vertical current sheets with associated changes in connectivity, i.e. the system demonstrates the mechanism of oscillatory reconnection.

% the fast magnetoacoustic wave develops into a fast oblique magnetic shock wave and significantly deforms the local magnetic fieldlines (specifically refracting the magnetic field away from the normal). The shocks deform the magnetic topology so much that the separatrices now touch one another rather than intersecting at a non-zero angle. The deformed null point continues to evolve and (locally) collapses to form a  horizontal current sheet.

%We also find that the magnetic shocks overlap and produce hot jets which substantially heat the local plasma (in our simulations, this occurs to the left and right of the equilibrium X-point). This localised heating plays a key role and a feedback effect occurs: the (local) plasma expands, squashing and shortening the horizontal current sheet, forcing the separatrices apart. The horizontal current sheet thus returns to a cusp-like null point which, due to the continuing expansion of the heated plasma, overshoots the equilibrium point, and in turn forms a vertical current sheet. The evolution then proceeds as before and we observe a clear oscillatory  cycle of horizontal and vertical current sheets. McLaughlin et al. (\cite{McLaughlin2009}) also present evidence for reconnection in their system (results we reproduce, but do not present here) and so the system was deemed to display the phenomenon of {\emph{oscillatory reconnection}}.

The main focus of this paper is on the periodic nature of this oscillatory cycle of  horizontal and vertical current sheets. For the first time, we identify two distinct phases in the oscillation: a transient, {\emph{impulsive phase}}, encompassing the development of the first horizontal current sheet, the formation of the first vertical current sheet, and ending with the formation of the second horizontal current sheet. We define the {\emph{stationary phase}} to begin  at the formation of the second horizontal current sheet and thus this phase includes all the proceeding cyclic behaviour.

%label the phase after the formation of the second horizontal current sheet, and all the proceeding cyclic behaviour, the {\emph{stationary phase}}.

In the impulsive phase, we find that {\emph{greater}} the driving amplitude ($C{\rm{v}}_0$) of the velocity initial condition (equation \ref{ICs}), [a] the {\emph{earlier}} the first horizontal current forms, [b] the {\emph{longer}} its maximum length and [c] the  {\emph{greater}} its maximum current density. These results are intuitive since we would expect magnetic shocks with larger amplitudes to propagate faster and thus arrive at the null point more rapidly than those with smaller amplitudes. We would also expect magnetic shocks with larger amplitudes to deform the pre-existing magnetic field to a greater extent (specifically to refract ${\bf{B}}$ away from the normal to a greater extent) and thus, ultimately,  to form {\emph{longer}} and {\emph{stronger}} current sheets.

We also investigate the time taken to evolve from the first horizontal current sheet to the second, which we labelled as the {\emph{impulsive period}}. We find that the  {\emph{greater}} the initial velocity amplitude ($C{\rm{v}}_0$) the {\emph{shorter}} the resultant impulsive period. Coupled with the results on current sheet length, this means that {\emph{longer current sheets have shorter corresponding impulsive periods}}. In this way, the system is acts as a {\emph{harmonic oscillator}}, i.e. the greater the displacement away from equilibrium, the stronger the restoring force, and thus the shorter the impulsive period.

For a driving amplitude of $25.2\:$km/s (corresponding to $C=1$ simulation) we measure an impulsive period of $54.3\:$s. We also investigate the resultant impulsive periods for driving amplitudes  $6.3 - 126.2\:$km/s  and find associated impulsive periods in the range $46.9 - 65.4\:$s.

%%%%%%%%%%%%%%%%%%%%%%%%%%%%%%%%%%%%%%%%%%%%

The stationary phase is found to be dominated by an exponentially-decaying oscillation, tending to a finite value, $j_{{\rm{final}}}$. We also investigate the  {\emph{stationary period}}, namely the time taken to evolve from the second horizontal current sheet to the third, i.e. the first complete cycle of the stationary phase.  As in the impulsive phase, we find that the {\emph{greater}} the initial velocity amplitude    ($C{\rm{v}}_0$) the {\emph{shorter}} the resultant stationary period. This is explained just as before, the {\emph{greater}} the initial amplitude, the longer and stronger the  current sheets at each stage, and thus the greater restoring force, leading to shorter periods (compared to smaller initial amplitude, shorter resultant current sheets, weaker restoring force and thus longer periods). Hence, again the system acts as a harmonic oscillator and, coupled with the exponential decay, we relate  the oscillatory reconnection mechanism to that of a {{damped harmonic oscillator}} during the stationary phase.

For a driving amplitude of $25.2\:$km/s (corresponding to $C=1$ simulation) we measure a stationary period of $69.0\:$s. We also investigate the resultant stationary periods for driving amplitudes  $6.3 - 126.2\:$km/s  and find associated stationary periods in the range $56.3 - 78.9\:$s, i.e. these are high frequency ($0.0127 - 0.0178\:$Hz) oscillations.

%%%%%%%%%%%%%%%%%%%%%%%%%%%%%%%%%%%%%%%%%%%%

{{

It is also prudent at this stage to ask what determines this stationary period and what determines the exponentially-decaying timescale. To this end, we can consider the work of Craig \& McClymont (\cite{CraigMcClymont1991}) who investigated the relaxation of a 2D X-point disturbed from equilibrium. By neglecting both nonlinear and thermal pressure effects, Craig \& McClymont (\cite{CraigMcClymont1991}) derived an analytical prediction for two timescales:
\begin{eqnarray}
\qquad t_{\rm{oscillation}} \approx   2 \ln {R_m}\;,\quad t_{\rm{decay}} \approx  t_{\rm{oscillation}}^2/ 2 \pi^2 \nonumber\;
\end{eqnarray}
where we identify $t_{\rm{oscillation}}$ as  our stationary period and $t_{\rm{decay}}$ as our decay time; $1/\lambda$.  For a  driving amplitude of $25.2\:$km/s, these correspond to  $t_{\rm{oscillation}} \approx 109.6\:$s, compared to our measured stationary period  of  $69.0\:$s, and $t_{\rm{decay}} \approx 76.7\:$s compared to our measured decay time of  $1/\lambda = 1/0.015 = 66.7\:$s, given that in our investigations $R_m= 10^4$ and time is made dimensional using $t_0=7.93\:$s. Thus, given the (relative) simplicity of the  Craig \& McClymont (\cite{CraigMcClymont1991}) system, these estimates are in fair agreement with our results. Note, however, that these simple analytical formulae cannot predict the variation in period versus amplitude of the initial velocity driver. This suggests that nonlinear effects and thermal-pressure gradients play a crucial role, which seems reasonable given that  the restoring force of oscillatory reconnection has been shown to be a dynamic  competition  between the thermal-pressure gradients and the Lorentz force (see $\S 3.2$ of Murray et al. \cite{Murray2009}; $\S 3.3$ of McLaughlin et al. \cite{McLaughlin2009};   Figure 7 of Threlfall et al. \cite{Threlfall2012}).

}}

 %and we suspect that the background environment is being (significantly) changed by the dynamic competition of forces at work behind oscillatory reconnection, where the restoring force of oscillatory reconnection is a time-dependent  competition  between the thermal-pressure gradients and the Lorentz force (see $\S 3.2$ of Murray et al. \cite{Murray2009}; $\S 3.3$ of McLaughlin et al. \cite{McLaughlin2009};   Figure 7 of Threlfall et al. \cite{Threlfall2012}).

%%%%%%%%%%%%%%%%%%%%%%%%%%%%%%%%%%%%%%%%%%%%

%It is interesting to note that the driving amplitudes and periods reported here are comparable to those reported in the obervational literature (e.g. Okamoto \& De Pontieu et al. \cite{OBart2011} report on transverse oscillations in spicules and 
% alfv\'en/tansversal waves chromospheric waves with amplitudes $10 - 30\:$km/s

%However, it should also be noted   => It is important to note that there is a difference between the impulsive period and the stationary period, and in every case the stationary period is longer than the corresponding impulsive period.  This indicates that different physical processes dominate in each phase, and thus validate our approach of dividing the whole evolution into two distinct phases.

It is also important to note that there is a significant difference between the impulsive period (e.g. $54.3\:$s for $C=1$ system) and the stationary period (e.g. $69.0\:$s for the $C=1$ system), and that in every numerical experiment we find that the stationary period is longer than the impulsive period. This indicates  that different physical processes dominate in each phase (i.e. the deformation of the X-point by the shock and the importance of jet heating in the impulsive phase, and the elastic motion of the magnetic field trying to get back to equilibrium in the stationary phase) and validates our approach of dividing the whole evolution into two distinct phases.

This difference in impulsive and stationary periodicities actually has an intriguing caveat : it is  important to note that when oscillatory reconnection is seen in, say, a numerical simulation, one must be careful to interpret  which phase one is actually observing {\emph{and}} to observe several oscillations, i.e. if only two periods are seen, say the impulsive period followed by a single stationary period, then one would conclude that the period was actually {\emph{increasing}} between oscillations. A similar result would pertain in solar observations of oscillatory reconnection, i.e. the first period measured would be shorter than the proceding periods (assuming the first, i.e. impulsive, period is also observed). This is a clear prediction for the oscillatory reconnection mechanism.

It is also important to note that, as shown by McLaughlin et al. (\cite{McLaughlin2012}), the  mechanism periodically generates ${\rm{v}}_x$ and  ${\rm{v}}_y$ but that these are {\emph{generated}} exponentially damped. Thus, if such signals are detected, then they may be decaying not due to a particular damping mechanism, but {\emph{due to the generation mechanism itself}}.

In addition to the stationary period, we also measured all the proceeding periods in the stationary phase, e.g. time taken to evolve from third horizontal current sheet to fourth, etc. Interestingly, it was found that the period very slightly decreases by roughly $1.8\%$ per oscillation. The exact reason for this decrease in the period is uncertain and may be a numerical effect. This will be investigated further in future work.

As in McLaughlin et al. (\cite{McLaughlin2009}), it was found that the final state (i.e. velocity zero, oscillatory behaviour ceased) is in force balance but is non-potential and a small, finite amount of current density exists in the system,  $j_{{\rm{final}}}$. The (final) X-point is very slightly closed up in the vertical direction, i.e. $j_z>0$ is associated with vertical current sheets. This is because the (local) plasma to the left and right of the X-point is slightly hotter, since that is where the initial, strongest, jet heating occurs. Thus, the existence of this  thermal-pressure gradient in force balance requires the final state to be non-potential. We find that the {\emph{greater}} the initial velocity amplitude   ($C{\rm{v}}_0$)        the {\emph{larger}} the value of $j_{{\rm{final}}}$. Again, this is intuitive: fast oblique magnetic shocks with a {\emph{greater}} amplitude will overlap  to form {\emph{stronger}}, {\emph{hotter}} jets to the left and right of the (equilibrium) X-point. Thus, this local plasma will be hotter at the end of the simulation, indicating a stronger thermal-pressure gradient and thus, in order to achieve force balance, a greater value of the Lorentz force, i.e. a {{larger}} value of $j_{{\rm{final}}}$.

We have presented an investigation into the periodic nature of oscillatory reconnection and have found that an aperiodic driver can {\emph{naturally}} generate a period signal via the physical mechanism of oscillatory reconnection. We have found that the system behaves akin to a {\bf{damped harmonic oscillator}}. Again, this is not surprising: effectively our velocity initial condition can be thought of as injecting a {\emph{finite amount of energy}} into the oscillatory reconnection mechanism and so intuitively the resultant periodic behaviour must be {\emph{finite in duration}}, i.e. this is a {\emph{dynamic}} reconnection phenomena as opposed to the classical steady-state, time-independent reconnection models.

Oscillatory Reconnection  may also play a role in generating  quasi-periodic pulsations  (see, e.g.,  reviews by Aschwanden \cite{Aschwanden2003}; Nakariakov \& Melnikov \cite{Nakariakov2009}). Oscillatory behavior has been reported  in a number of solar and stellar flare observations (e.g. Mathioudakis et al. \cite{Mathioudakis2003}; \cite{Mathioudakis2006}; McAteer et al. \cite{McAteer2005}; Inglis et al. \cite{Inglis2008}; Inglis \& Nakariakov \cite{Inglis2009}; Nakariakov et al. \cite{Nakariakov2010};  Nakariakov \& Zimovets \cite{Nakariakov2011}; Inglis \& Dennis \cite{Inglis2012}; Shen \& Liu \cite{Shen2012}) but the generation mechanism responsible remains an open question.

We believe the physical mechanism of oscillatory reconnection described in this paper is a robust, general phenomenon that will be observed in other systems that demonstrate finite-duration/non-steady-state reconnection (although we have only presented a specific example of oscillatory reconnection in this paper).  For example, evidence of oscillatory reconnection in 3D flux emergence simulations has been  reported by Archontis et al. (\cite{VasilisOSCILLATORYRECONNECTION}).

%----------------------------------------------------------------
%----------------------------------------------------------------
%----------------------------------------------------------------
%----------------------------------------------------------------
%----------------------------------------------------------------
\begin{acknowledgements}
{The authors acknowledge IDL support provided by STFC. JAM and JOT acknowledge financial assistance from the Royal Astronomical Society. The computational work for this paper was carried out on the joint STFC and SFC (SRIF) funded cluster at the University of St Andrews (Scotland, UK).}
\end{acknowledgements}
%----------------------------------------------------------------
%----------------------------------------------------------------
%----------------------------------------------------------------
%----------------------------------------------------------------
%----------------------------------------------------------------

%----------------------------------------------------------------
%----------------------------------------------------------------
%----------------------------------------------------------------
%----------------------------------------------------------------
%----------------------------------------------------------------
%

%----------------------------------------------------------------
%----------------------------------------------------------------
%----------------------------------------------------------------
%----------------------------------------------------------------
%----------------------------------------------------------------

\end{document}